\documentclass[
reprint,
superscriptaddress,
showpacs,
 amsmath,amssymb,
floatfix,
]{revtex4-1}
\usepackage{dcolumn}
\usepackage{bm}
\usepackage{turnstile}

\usepackage{graphicx}
\usepackage{amsmath,amssymb}
\usepackage{url}
\usepackage{multirow}
\usepackage{float}

\DeclareMathOperator{\sgn}{\mathop{\mathrm{sgn}}}
\DeclareMathOperator{\re}{\mathop{\mathrm{Re}}}

\newcommand{\Eq}[1]{Eq.~(\ref{#1})}
\newcommand{\Eqs}[1]{Eqs.~(\ref{#1})}
\begin{document}

\title{
Magnetic and Superconducting Phase Diagram of Nb/Gd/Nb trilayers
}
\author{Yu.~N.~Khaydukov}
\affiliation{Max-Planck-Institut f\"ur Festk\"orperforschung, Heisenbergstra\ss e 1, D-70569 Stuttgart, Germany}
\affiliation{Max Planck Society Outstation at the Heinz Maier-Leibnitz Zentrum (MLZ), D-85748 Garching, Germany}
\affiliation{Skobeltsyn Institute of Nuclear Physics of Moscow State University, 119991 Moscow, Russia}

\author{A.~S.~Vasenko}
\affiliation{National Research University Higher School of Economics, 101000 Moscow, Russia}

\author{E.~A.~Kravtsov}
\affiliation{Institute of Metal Physics, 620180 Ekaterinburg, Russia}
\affiliation{Ural Federal University, 620002 Ekaterinburg, Russia}
\author{V.~V.~Progliado}
\affiliation{Institute of Metal Physics, 620180 Ekaterinburg, Russia}

\author{V.~D.~Zhaketov}
\affiliation{Joint Institute for Nuclear Research, 141980 Dubna, Russia}
\author{A.~Csik}
\affiliation{Institute for Nuclear Research (Atomki) Hungarian Academy of Sciences, Debrecen, Hungary}
\author{Yu.~V.~Nikitenko}
\affiliation{Joint Institute for Nuclear Research, 141980 Dubna, Russia}
\author{A.~V.~Petrenko}
\affiliation{Joint Institute for Nuclear Research, 141980 Dubna, Russia}

\author{T.~Keller}
\affiliation{Max-Planck-Institut f\"ur Festk\"orperforschung, Heisenbergstra\ss e 1, D-70569 Stuttgart, Germany}
\affiliation{Max Planck Society Outstation at the Heinz Maier-Leibnitz Zentrum (MLZ), D-85748 Garching, Germany}

\author{A.~A.~Golubov}
\affiliation{Faculty of Science and Technology and MESA$^+$ Institute for Nanotechnology,
University of Twente, 7500 AE Enschede, The Netherlands}
\affiliation{Moscow Institute of Physics and Technology, Dolgoprudny, 141700 Moscow, Russia}

\author{M.~Yu.~Kupriyanov}
\affiliation{Skobeltsyn Institute of Nuclear Physics of Moscow State University, 119991 Moscow, Russia}

\author{V.~V.~Ustinov}
\affiliation{Institute of Metal Physics, 620180 Ekaterinburg, Russia}

\author{V.~L.~Aksenov}
\affiliation{Joint Institute for Nuclear Research, 141980 Dubna, Russia}
\author{B.~Keimer}
\affiliation{Max-Planck-Institut f\"ur Festk\"orperforschung, Heisenbergstra\ss e 1, D-70569 Stuttgart, Germany}
\date{\today}

\begin{abstract}
We report on a study of the structural, magnetic and superconducting properties of Nb(25nm)/Gd($d_f$)/Nb(25nm) hybrid structures of a superconductor/ ferromagnet (S/F) type. The structural characterization of the samples, including careful determination of the layer thickness, was performed using neutron and X-ray scattering with the aid of depth sensitive mass-spectrometry. The magnetization of the samples was determined by SQUID magnetometry and polarized neutron reflectometry and the presence of magnetic ordering for all samples down to the thinnest Gd(0.8nm) layer was shown. The analysis of the neutron spin asymmetry allowed us to prove the absence of magnetically dead layers in junctions with Gd interlayer thickness larger than one monolayer. The measured dependence of the superconducting transition temperature $T_c(d_f)$ has a damped oscillatory behavior with well defined positions of the minimum at $d_f$=3nm and the following maximum at $d_f$=4nm; the behavior, which is in qualitative agreement with the prior work (J.S. Jiang et al, PRB 54, 6119). The analysis of the $T_c(d_f)$ dependence based on Usadel equations showed that the observed minimum at $d_f$=3nm can be described by the so called "$0$" to "$\pi$" phase transition of highly transparent S/F interfaces with the superconducting correlation length $\xi_f \approx 4$nm in Gd. This penetration length is several times higher than for strong ferromagnets like Fe, Co or Ni, simplifying thus preparation of S/F structures with $d_f \sim \xi_f$ which are of topical interest in superconducting spintronics.

\end{abstract}


\maketitle

\section{Introduction}

Superconductor/ ferromagnet (S/F) hybrid structures are attracting great interest nowadays due to a large number of phenomena, including $\pi$ Josephson junctions, non-monotonous dependence of the critical temperature $T_c$ on the thickness of the F layer $d_f$, superconducting spin-valves, triplet superconductivity, etc \cite{BuzdinRevModPhys,GolubovRMP,BergeretRMP,eschrig2011,SidorenkoLTP17}. This rich physics is based on the proximity effect - i.e. the penetration of superconducting correlations from the S into the F layer over the typical distance  $\xi_f$ of order of 1-10 nm.  This leakage leads to the damped oscillatory behavior of the pairing potential in S/F multilayers. Even for the simplest system, i.e. a S/F bilayer, this effect leads to a non-trivial $T_c$($d_f$) dependence: depending on the interface transparency the $T_c$($d_f$) function can be oscillating, re-entrant for highly transparent interfaces or monotonously decaying for interfaces with medium or low transparency \cite{Fominov.PRB2002,ZdravkovPRL,Zdravkov.PRB,AttanasioPRB2005,AttanasioPRB2007}. For a larger number of S/F interfaces the behavior of the pairing potential becomes more complicated. For $d_f < \xi_f$, the pair wave function in the F layer changes little and the superconducting pair potential in the adjacent S layers remains the same. The phase difference between the pair potentials in the S layers is then absent, which is referred to as the "0" phase state. On the other hand, if $d_f \sim \xi_f$, the pair wave function may cross zero at the center of the F layer with an opposite sign or $\pi$ shift of the phase of the pair potential in the adjacent S layers, which is called the "$\pi$" phase state \cite{buzdin1992zhetf,Demler1997}. An increase of the F layer thickness $d_f$ may provoke subsequent transitions from 0 to $\pi$ phases or even into more complex phases \cite{KushnirPRB2011,KushnirJETPlett2011}. The existence of "$\pi$" state leads to a number of striking phenomena. For example, the critical current in S/F/S Josephson junctions exhibits a damped oscillatory behavior with increasing F layer thickness \cite{buzdin1991josephson,Ryazanov2001, Oboznov2006, Weides2006, Kemmler2010}. In the $\pi$ state the critical current is negative, and the transition from the 0 to the $\pi$ state results in a sign change of the critical current. Zero to $\pi$ transitions can be also observed as density of states (DOS) oscillations \cite{Kontos2001, Buzdin2000, Vasenko2008, Vasenko2011}, critical temperature $T_c$ oscillations \cite{JiangPRL,JiangPRB} or peculiarities in electrodynamics \cite{SilvaPhysRevB14} of S/F multilayers.

The S/F structures attract interest not only from the scientific but also from the technological point of view as elements of superconducting spintronics \cite{soloviev2017beyond,GolubovNatMat17,SidorenkoLTP17,AartPhysRevX15,eschrig2011,FominovJetpLett2010,TagirovPRL1999}. High performance of such devices is predicted and realized for highly transparent S/F interfaces with $d_f \sim \xi_f$.

One of the first S/F systems which was proved to have high transparency were Gd/Nb systems \cite{JiangPRL,JiangPRB}. A series of Nb/Gd/Nb trilayers and periodic structures were prepared using magnetron sputtering, and an oscillatory  $T_c(d_f)$ behavior was observed. More  recently,  homogeneous NbGd alloys \cite{bawa2016(alloy)} and GdN/Nb/GdN trilayers \cite{BritScience17} were studied. However pure gadolinium in combination with niobium has several advantages compared to other S/F systems widely used nowadays.  First, gadolinium is a  localized ferromagnet with rather low (compared to Fe,Co and Ni) bulk Curie temperature of $T_m$ = 292 K \cite{Koehler1972}. Strong localization of the magnetic moment stabilizes ferromagnetism even in ultra-thin Gd layers. In contrast itinerant ferromagnets (Fe, Co, Ni) form magnetically dead layers at the S/F interface \cite{Muege97,Obi1999,ObiPRL}, thus deteriorating the interface transparency. Another advantage of Gd is its ability to couple with other ferromagnets \cite{Choi2004(GdFe),Kravtsov2009,Ryabukhina2015,SanyalPRL104,higgs2016} forming non-trivial magnetic ordering patterns which can be used for the creation of superconducting spin-valves \cite{FominovJetpLett2010,AartPhysRevX15,TagirovPRL1999}. Finally, niobium and gadolinium components are not mutually soluble, neither in the solid nor in the liquid phase \cite{Elliot1965,Shunk1969}. Motivated by these arguments we have prepared and thoroughly studied a series of Nb(25nm)/Gd($d_{f}$)/Nb(25nm) trilayers. Our work complements and expands the pioneering work of Jiang et al \cite{JiangPRB}.

\section{Sample fabrication and experimental techniques}

The samples of nominal structure  Ta(3nm)/Cu(4nm)/Nb(25nm)/Gd($d_f$)/Nb(25nm) (here and later SFSx, where $x\equiv d_f$ measured in nanometers) were prepared using an UHV magnetron machine ULVAC MPS-4000-C6 at constant current onto Al$_2$O$_3$(1$\bar{1}$02) substrates with thickness of Gd layer $d_f = [0.8 \div 7.5]$nm (see inset to Fig. \ref{Fig1}). The bilayer Ta/Cu on the top is required to protect against oxidation and to create a neutron waveguide structure \cite{khaydukov2015peculiarities}.

Before the deposition, the substrate was cleaned from organic contaminations with acetone and alcohol. The substrate was further cleaned in-situ with reverse magnetron sputtering (2 minutes at an argon flow rate of 25 sccm) in the load chamber. The base pressure was lower than $2\times10^{-9}$ mbar. Pure argon gas (99.9998\% purity) at a flow rate of 25 sccm was used as sputter gas. The deposition was carried out at room temperature (about 25 $^\circ$C) at a magnetron sputtering power of 100 Watts in an argon atmosphere of $1\times10^{-3}$ mbar. In these conditions Nb, Gd, Cu, and Ta layers were sputtered at deposition rates of 2.35 nm/min, 6.85 nm/min, 6.45nm/min, and 2.8 nm/min, respectively. The deposition rates were calibrated using test samples with the help of a Zygo NewView7300 white light interferometer.

The quality of the layers and interfaces was studied by  Secondary Neutral Mass Spectrometry (SNMS, SPECS GmbH Berlin INA-X type), X-ray and neutron reflectometry. Both X-ray and neutron reflectometry allows one to reconstruct the depth profile of X-ray/neutron scattering length density (SLD) \cite{XNrefl2008}. In addition, the SLD of neutrons is spin-dependent: $\rho^{\pm}=\rho_0(z) \pm cM(z) $, where the superscript denotes the sign of the neutron spin projection on the external field, $\rho_0(z)$ and M(z) - are the depth profiles of the nuclear SLD and the in-plane magnetization, and c = 0.231 $\times 10^{-4}$ nm$^{-2}$/kG is a scaling factor. Polarized Neutron Reflectometry (PNR) can thus be used as a depth sensitive magnetometric method. In the part of the PNR measurements at remanence we measured intensity of spin-flip scattering. This scattering channel allowed us to obtain information about the component of the in-plane magnetization non-collinear to the external field. We note that PNR is sensitive only to the in-plane component of the magnetization. The X-ray reflectivity curves were measured on the PANalytical Empyrean diffractometer at wavelength $\lambda$ = 0.229 nm. In addition to PNR we used SQUID magnetometry for the magnetic measurements. The PNR experiments were conducted on the angle-dispersive reflectometer NREX ($\lambda$ = 0.428 nm) at the research reactor FRM-II (Garching, Germany) and Time-of-Flight reflectometer REMUR ($\lambda = [0.15 \div 1]$ nm) at the research reactor IBR-2 (Dubna, Russia). In all magnetometric measurements, the external magnetic field was applied in-plane of the structure. Superconducting properties were measured by a SQUID magnetometer and a mutual inductance setup.

\section{Structural properties}
\begin{figure*}[htb]
\centering
\includegraphics[width=2\columnwidth]{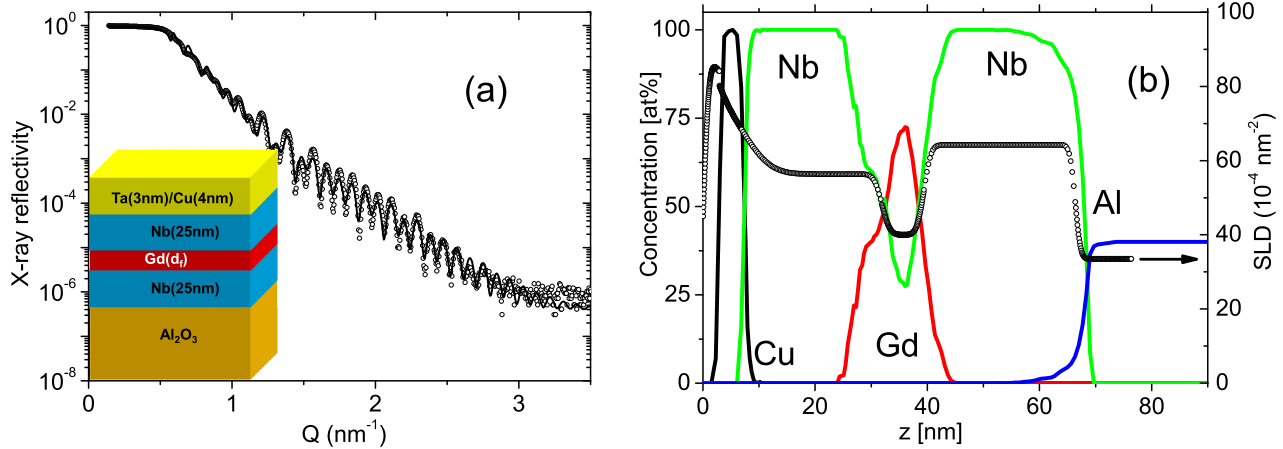}
\caption{
(a) The X-ray experimental (dots) and theoretical (solid curves) reflectivity curves for the sample SFS7.5. The inset shows the general design of the prepared structures. (b) Concentration profiles of Cu, Nb, Gd and Al elements for the same sample measured by SNMS. The dots show the X-ray SLD depth profile for the model reflectivity depicted in (a).
}
\label{Fig1}
\end{figure*}

A typical X-ray reflectivity curve measured on the sample SFS7.5 is shown in  Fig. \ref{Fig1}a. The curve exhibits so-called Kiessig oscillations caused by the interference of X-rays reflected from different interfaces inside the structure. The experimental curve can be reasonably well fitted by the model reflectivity calculated for the SLD depth profile depicted in Fig. \ref{Fig1}b. In the same figure we show the concentration depth profile measured by SNMS. One can see a good correspondence of the layer thickness obtained by different methods. By further analysis of the depth profiles we can conclude that the real thickness of the layers deviates by at most 10\% from the nominal values and that the interfaces are characterized by an rms roughness of order 1nm.

Similar data treatment was performed for the other samples. In the cases when SNMS was not measured, we fitted X-ray and neutron reflectivity curves simultaneously keeping the parameters $d_s$ and  $d_f$ the same for both curves. Thicknesses of all samples are within 10 \% of the nominal values. The rms roughness of the Gd/Nb interfaces obtained from the fits is 0.5-1 nm.

\section{Magnetic properties}
\begin{figure}[htb]
\centering
\includegraphics[width=\columnwidth]{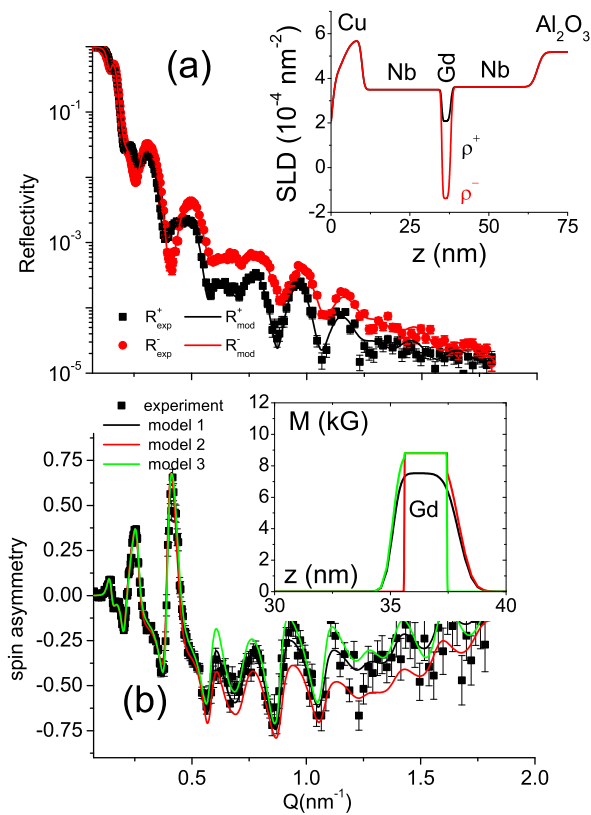}
\caption{
(Color online). (a) Experimental (dots) and model (solid line) neutron reflectivity curves for the sample SFS3 measured at T = 6.2K and  H = 4kOe. Inset shows the SLD depth profiles for spin-up ($\rho^+$) and spin-down ($\rho^-$) neutrons corresponding to the best-fit model.  (b) The experimental (dots) spin asymmetry for the reflectivities depicted in the panel (a). The solid lines show the model curves for the magnetic profiles which are shown in the inset.
}
\label{Fig2}
\end{figure}

The spin-polarized neutron reflectivities measured on the sample SFS3 at $T$=6.2K and $H=$ 4kOe are shown in Fig. \ref{Fig2}a. The non-zero spin asymmetry $S \equiv (R^+ - R^-)/(R^+ + R^-)$ (Fig. \ref{Fig2}b) evidences the presence of a magnetic moment in our system. The inset to Fig.\ref{Fig2}a shows the depth profiles $\rho^+(z)$ and $\rho^-(z)$ corresponding to the best-fit model. One can see that the splitting of the curves is due to the presence of a magnetization M = 7.5kG in the Gd layers (here and later we assume  that the magnetization is already multiplied by 4$\pi$ factor).

We paid particular attention to possible magnetic dead layers in our samples. First of all we note that the sample with thinnest $d_f$ = 0.8 nm is still ferromagnetic, which gives us a lower bound on the thickness $d_{DL}$ of the dead layer. We also included dead layers in our models of the PNR data. In Fig. \ref{Fig2}b we show the calculated spin asymmetries for 3 models: no magnetic dead layer (model 1), a dead layer with thickness $d_{DL}$ = 0.5nm at the bottom S/F interface (model 2) and a dead layer at the top S/F interface (model 3). In all models  the total magnetic moment is constrained to be equal to the macroscopic moment measured by SQUID magnetometry. The model 1 provides the best description of the data with goodness of the fit $\chi^2 $ = 8.6. The models 2 and 3 show worse agreement with experiment with  $\chi^2 $ = 9.8 and  $\chi^2 $ = 9.5, correspondingly. We also tried to model the presence of dead layers on both interfaces and end up with  $\chi^2 $ = 9.7.

\begin{figure*}[!ht]
\centering
\includegraphics[width=2\columnwidth]{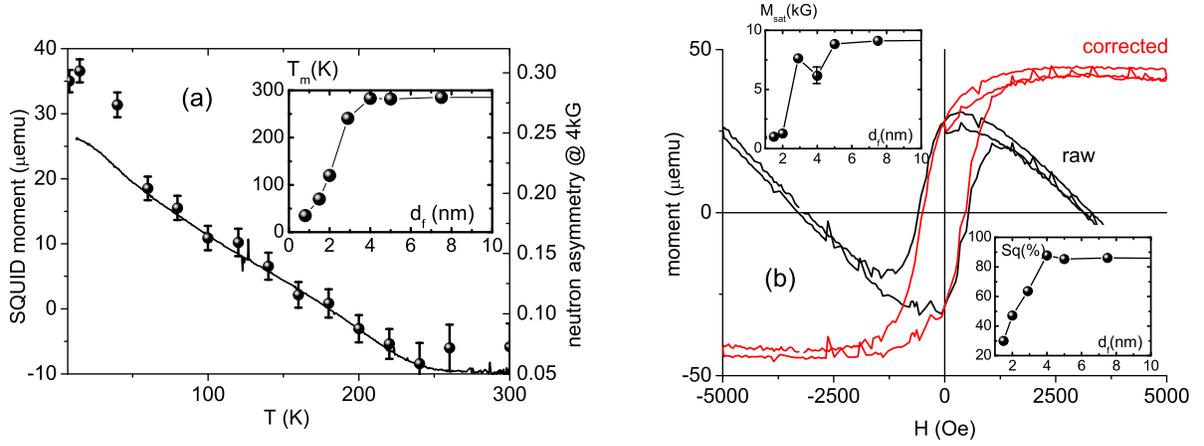}
\caption{
(Color online)(a) The temperature dependence of the SQUID magnetic moment (solid line) and averaged spin asymmetry (dots) for the sample SFS3. The inset shows the dependence of $T_m(d_f)$. (b) The field dependence for the sample SFS3  measured at T = 13K. Original and background corrected loops are shown by black and red curves, respectively. Upper and bottom insets show the $d_f$-dependencies of the saturation magnetization and the squareness of the hysteresis loop.
}
\label{Fig3}
\end{figure*}

The temperature dependence of the magnetic moment measured by SQUID in a magnetic field $H$ = 661 Oe on the sample  SFS3 is shown in Fig. \ref{Fig3}a. The SQUID measurements have to be carried out in low magnetic fields due to the diamagnetic response of the substrate (see Fig. \ref{Fig3}b). The smaller $d_f$, the higher the field range where the magnetic signal of the substrate dominates over the signal of the F layer.  The PNR data, in contrast, are insensitive to the magnetic moment of the substrate and can be measured in fields above saturation. In the same Fig. \ref{Fig3}a we show the temperature dependence of the neutron spin asymmetry  measured in applied field $H$ = 4 kOe. One can see a good agreement between neutron and SQUID data at all temperatures down to $T$=60K. The difference at lower temperature can be ascribed to a reorientation of the easy axis which was observed in bulk Gd \cite{bulkGd}. The Curie temperature $T_m$ was extracted from the temperature dependence of $m(T)$; and the resulting dependence $T_m(d_f)$ is shown in the inset to Fig. \ref{Fig3}a. The $T_m$ grows with increasing $d_f$ up to $d_f \sim$ 3nm and then saturates at the bulk value. This behavior is in qualitative agreement with previous reports \cite{JiangPRL}.

The field dependence of the SQUID magnetic moment measured at T=13K on the same sample is shown in Fig. \ref{Fig3}b. The hysteresis loop reveals a coercivity field of $H_c \approx$ 500 Oe and saturation magnetic moment of $m_{sat} \approx$ 50 $\mu$emu. Knowing from XRR and NR the Gd layer thickness and sample area $S$ = 25 mm$^2$ the saturation magnetization can be calculated as $M_{sat} = m_{sat}/(d_f \times S$)=7.6kG. This value is in good agreement with M=7.5kG found from PNR, giving thus another cross-check of our determination of the thickness $d_f$ and magnetization. Upper inset in Fig. \ref{Fig3}b shows the dependence $M_{sat}(d_f)$. One can also see that $M_{sat}(d_f)$ correlates with $T_m(d_f)$ depicted in the inset to Fig. \ref{Fig3}a.

Another characteristic of a hysteresis loop is its squareness $Sq\equiv m_{rem}/m_{sat} $ ($m_{rem} $ is the remanent magnetic moment).  A squareness less than 100\% means that the external magnetic field was applied at an angle $\alpha \approx acos(Sq)$ to the easy axis direction.  The $Sq(d_f)$ dependence shown in the bottom inset to Fig. \ref{Fig3}b tells us that the easy axis (EA) of all samples makes a non-zero angle with $H$. If the EA lies in-plane of the structure we should observe neutron spin-flip scattering, which, however, was not observed in the experiment. This leads us to the conclusion that the EA is aligned out of plane.

\section{Superconducting properties and proximity effect}
\begin{figure*}[!ht]
\centering
\includegraphics[width=2\columnwidth]{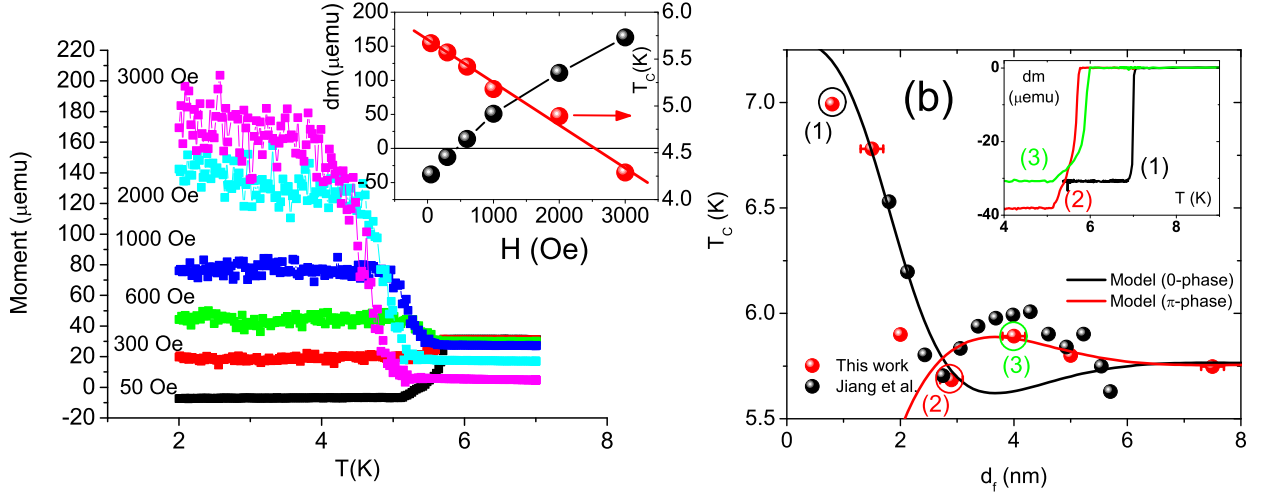}
\caption
 {
 (Color online). (a) Temperature dependence of the magnetic moment measured by SQUID on the sample SFS3. The inset shows the field dependence of the upturn amplitude and $T_c$ (defined as the midpoint of the transition). (b) Experimental $T_c(d_f)$  dependence for this work (red dots) and Ref.~\cite{JiangPRB} (black dots) with $\delta d_f =$ 1.5nm correction. Black and red lines show calculations using the Usadel equations for 0 and $\pi$ phase states. The inset shows the temperature dependence of the magnetic moment change $ dm(T) \equiv m(T)-m(T>T_c)$ for the set of samples marked in the main panel.
}
\label{Fig4}
\end{figure*}

The temperature dependencies of magnetic moment around $T_c$ in different magnetic fields measured by SQUID on the SFS3 sample are shown in Fig. \ref{Fig4}a. In magnetic fields $H < 0.5$ kOe the total magnetic moment decreases below $T_c$ due to the Meissner response of the S layers. For  fields $H > 0.5$ kOe, in contrast, an increase of the magnetic moment is observed with almost linear dependence of the jump on magnetic field (inset to Fig. \ref{Fig4}a). A similar jump with linear dependence was observed recently in (Fe,Co,Ni)/V bilayers \cite{NagyEPL}. It is interesting to note that PNR in contrast does not reveal any difference of the spin asymmetries above and below $T_c$ within the statistical accuracy.

The field dependence of $T_c$ is shown by red dots in the inset to Fig. \ref{Fig4}a. The $T_c(H)$ dependence for this sample has a linear form typical for 3D superconductors which evidences coupling of the S layers through the F one. By fitting the experimental dependence $T_c(H)$ we can extract $T_c(0)$=5.7K and $H_{c2}(0)$=12.6kOe. The latter value allows us to estimate the superconducting correlation length  $\xi_s \sim$ 10 nm.

The $T_c(d_f)$ dependence is shown in Fig. \ref{Fig4}b. It has a damped oscillatory behavior with a minimum at $d_f$ = 3nm followed by a maximum at $d_f$ = 4nm. The shape of the curve is similar to the one measured by Jiang et al \cite{JiangPRB}.  for similar Nb(25nm)/Gd/Nb(25nm) trilayers, however both minimum and maximum are observed in our case at $\delta d_f \sim$ 1.5nm higher values. After taking this offset into account, both $T_c(d_f)$ dependencies become consistent.

\section{Discussion and conclusion}
In this work we have studied the $d_f$-dependence of the magnetic and superconducting properties of Nb(25nm)/Gd($d_f$)/Nb(25nm) S/F/S trilayers. The thickness $d_f$ was derived from comprehensive analysis of several structural and magnetic techniques such as Secondary Neutral Mass Spectrometry, X-ray and neutron reflectometry and SQUID magnetometry. The magnetic and superconducting transition temperatures generally agree with the ones reported in Ref.\cite{JiangPRL}, if an off-set in thickness of the Gd layer $\delta d_f \sim$ 1.5nm is taken into account. The difference can not be explained by the presence of magnetically dead layers in our structures, since both PNR and SQUID data exclude the presence of any dead layer with thickness more than one monolayer. Taking into account the similar Curie temperatures of the S/F/S systems in this work and analogous trilayers in Ref.~\cite{JiangPRB}  the off-set can not be explained by different experimental conditions either. We attribute the difference to a mis-calibration of the thicknesses of the Gd layer in Refs.~\cite{JiangPRB,JiangPRL}. In these works the thicknesses were calibrated using a quartz-crystal monitor and the position of the Bragg peaks in the X-ray reflectivities. However, the quartz-crystal monitor has a sensitivity of order of 1nm and the position of the low-order Bragg peaks are shifted towards higher angles due to the refraction effect. This means that attempts to calculate the thickness of a Nb/Gd bilayer using the standard Bragg law will generate a systematic error of 1-3nm (see Appendix B). Taking into account this off-set allows us to reconcile the dependencies of $T_c(d_f)$  in both cases.

We compared the experimental $T_c(d_f)$ dependencies to model curves calculated for $0$ and $\pi$ states using the Usadel approach (see Appendix A). For the calculations we fixed $\xi_s$=10nm and the exchange energy $E_{ex}$=280K and varied $\xi_f$, $\gamma$ and $\gamma_b$. The parameters $\gamma$ and $\gamma_b$ are expressed via the normal-state conductivity of the S(F) layer, $\sigma_{s(n)}$, the resistance of the S/F boundary, $R_B$, and above defined correlation lengths as $\gamma = \xi_s \sigma_n / \xi_n \sigma_s$, $\gamma_B = R_B \sigma_n / \xi_n$. Reasonably good agreement between experiment and theory was obtained for $\gamma$=0.07, $\gamma_b \to $0 and $\xi_f$ = 4 nm. The extremely small parameter $\gamma_b$ indicates a high transparency of the S/F interface. Thus, according to our calculations the superconducting correlations penetrate into Gd layer on a typical length of $\xi_f$ = 4 nm. The  $\pi$ state becomes energetically favorable for the region of thickness $d_f=[3 \div 6]$ nm. For higher thicknesses transmission of the correlations through the F layer becomes impossible and the S/F/S structure splits into two independent S/F bilayers.

We now discuss the magnetic properties of the samples. Our investigation has shown that the samples with $d_f>$ 2nm have Curie temperatures close to the bulk and almost square hysteresis loops. However, the magnetic moment of our structures is only 3.7 $\mu_B$/Gd i.e. roughly half of the bulk value. A similarly suppressed moment  was found in Gd/U \cite{GdU}, Gd/V \cite{GdV} and Gd/Cr \cite{GdCr} multilayers. This suppression may well be related to the presence of the less magnetic fcc phase together with the bulk hcp phase which was recently found in Fe/Cr/Gd multilayers \cite{Ryabukhina2015}.

Below $T_c$ we have observed an upturn of the magnetic moment if the sample was cooled down in certain magnetic field. A similar upturn, often called Paramagnetic Meissner Effect (PME) was already observed in several prior works \cite{SatapathyPRL,BernardoPME,TorrePME,MontonPME,OvsyannikovPME,NagyEPL} and explained either by electrodynamical or exchange coupling mechanisms. Based on (a) the observation of the effect at high fields, (b) the linear field dependence of the enhanced moment and (c)the absence of the effect in PNR we attribute the PME in our samples to out-of-plane vortices. In Ref. \cite{KoshelevPME} the PME for a single S film with external field directed normal to the surface was explained by vortex trapping. In our case the stray field of Gd can play the role of the out-of-plane external field. It is also known that the proximity effect can influence the vortex dynamics and hence cause a PME \cite{GolubovJETP16}. This question has to be addressed by future investigations.

In conclusion, we have shown that high quality Nb/Gd/Nb trilayers can be grown using magnetron sputtering in a wide range of thicknesses. The penetration depth of superconducting correlations in the Gd layer is found to be several times higher than for strong ferromagnets like Fe, Co or Ni. This simplifies preparation of S/F structures with $d_f \sim \xi_f$ which are of topical interest in superconducting spintronics.

\begin{acknowledgments}
The authors would like to thank G.Nowak and V. Zdravkov for fruitful discussion and R. Morari for the assistance in the experiment. This work is partially based on experiments performed at the NREX instrument operated by Max-Planck Society at the Heinz Maier-Leibnitz Zentrum (MLZ), Garching, Germany and supported by DFG collaborative research center TRR 80. Theoretical modeling of superconducting properties performed by A.S.V. was supported by joint Russian-Greek projects RFMEFI61717X0001 and T4$\Delta$P$\Omega$-00031 "Experimental and theoretical studies of physical properties of low-dimensional quantum nanoelectronic systems". The research in Ekaterinburg has been carried out within the state assignment  on  the  theme  "Spin"  No.  01201463330 and  with the support of the Ministry of Education and Science of the Russian Federation (Grant No. 14.Z50.31.0025). The X-ray measurements were performed at the Collective Use Center of the IMP. SNMS measurements were carried out in frame of the GINOP-2.3.2-15-2016-00041 project, which is co-financed by the European Union and the European Regional Development Fund.

\end{acknowledgments}


\appendix

\section{Critical temperature calculation}
\renewcommand{\theequation}{A\arabic{equation}}
\setcounter{equation}{0}

\begin{figure}[htb]
\centering
\includegraphics[width=\columnwidth]{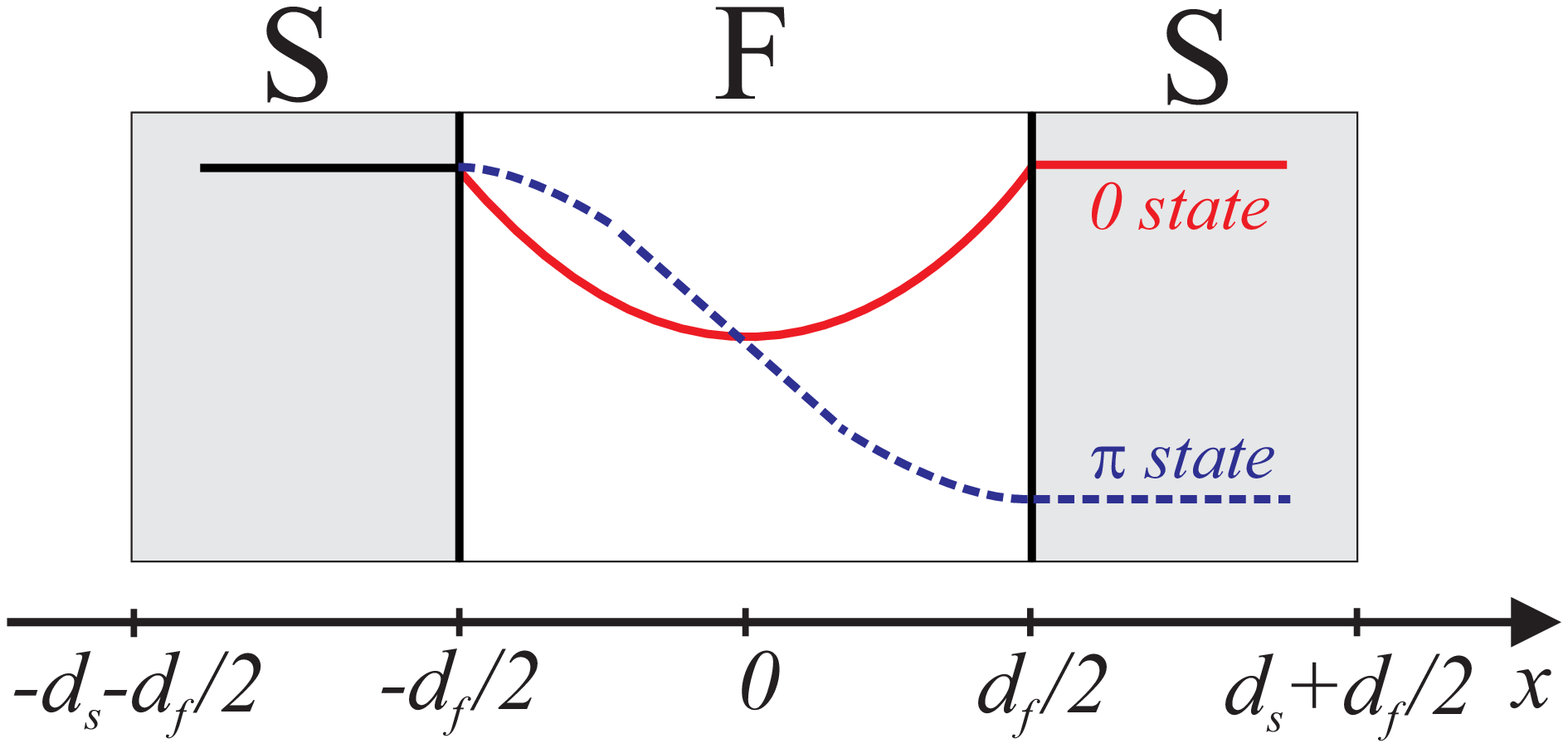}
\caption
{
(Color online). Geometry of the considered system. The thickness of the ferromagnetic interlayer is $d_f$. The typical behavior of the real part of the pair wave function $F$ is shown schematically. The pair wave function in zero state is shown by solid red line, while in $\pi$ state - by dashed blue line. Only one of these states is realized depending on the F layer thickness $d_f$.
}
\label{SIFS}
\end{figure}

The model of an S/F/S junction we are going to study is depicted in Fig.~\ref{SIFS} and consists of a ferromagnetic layer of
thickness $d_f$ and two superconducting layers of thickness $d_s$ along the $x$ direction. The structure is symmetric and its center is placed at $x=0$. We assume the diffusive limit and $\hbar = k_B = 1$.

To calculate the critical temperature $T_c(d_f)$ of this structure we use the framework of the linearized Usadel equations for the S and F layers. Near $T_c$ the normal Green's function is $G = \sgn \omega_n$, and the Usadel equations for the anomalous Green's function $F$ in the S layers reads ($d_f/2< |x| <d_s + d_f/2$) \cite{Usadel}
\begin{equation}\label{Usadel_S}
\xi_s^2 \pi T_{cs} \frac{d^2 F_s}{d x^2} - |\omega_n| F_s + \Delta = 0.
\end{equation}
In the F layer ($-d_f/2 < x < d_f/2$) the Usadel equation can be written as \cite{BuzdinRevModPhys},
\begin{equation}\label{Usadel_F}
\xi_n^2 \pi T_{cs} \frac{d^2 F_f}{d x^2} - \left ( |\omega_n| + i E_{ex} \sgn \omega_n \right ) F_f = 0.
\end{equation}
Finally, the selfconsistency equation reads \cite{BuzdinRevModPhys},
\begin{equation}\label{Delta}
\Delta \ln \frac{T_{cs}}{T} = \pi T \sum_{\omega_n} \left ( \frac{\Delta}{|\omega_n|} - F_s \right ).
\end{equation}
Here $\xi_s = \sqrt{D_s/ 2\pi T_{cs}}$, $\xi_n = \sqrt{D_f/ 2\pi T_{cs}}$, $\omega_n = 2 \pi T (n + \frac{1}{2})$ with $n = 0, \pm 1, \pm 2, \ldots$ are the Matsubara frequencies, $E_{ex}$ is the exchange field in the ferromagnet, $T_{cs}$ is the critical temperature of the S material ($d_f \to 0$), and $F_{s(f)}$ denotes the anomalous Green's function in the S(F) region. We note that $\xi_f = \xi_n \sqrt{2\pi T_{cs} / E_{ex}}$.

Equations \eqref{Usadel_S}-\eqref{Delta} must be supplemented by the following boundary conditions at the  S/F interfaces ($x=\pm d_f/2$) \cite{KL},
\begin{subequations}\label{KL}
\begin{align}
\xi_s \frac{d F_s(\pm d_f/2)}{dx} &= \gamma \xi_f \frac{d F_f(\pm d_f/2)}{dx},
\\
\xi_f \gamma_b \frac{d F_f(\pm d_f/2)}{dx} &= \pm F_s(\pm d_f/2) \mp F_f(\pm d_f/2),
\end{align}
\end{subequations}
where $\gamma = \xi_s \sigma_n / \xi_n \sigma_s$, $\sigma_{s(n)}$ is the normal-state conductivity of the S(F) layer, $\gamma_B = R_B \sigma_n / \xi_n$ \cite{KL, gamma_B1,gamma_B2}, and $R_B$ is the resistance of the S/F interfaces (we assume a symmetric structure with same resistance $R_B$ for $x=\pm  d_f/2$). At the borders of the S layer with a vacuum we naturally have,
\begin{equation}\label{S-vac}
\frac{d F_s( \pm d_s \pm d_f/2)}{dx} = 0.
\end{equation}

The solution of the Usadel equation in the F layer depends on the phase state of the structure. In the 0 phase state the anomalous Green's function is symmetric relative to $x=0$ (see Fig.~\ref{SIFS}) \cite{Fominov.PRB2002},
\begin{align}
F_f^0 &= C(\omega_n) \cosh \left( k_f x \right),\label{zero}
\\
\quad k_f &= \frac{1}{\xi_n}\sqrt{\frac{|\omega_n| + i E_{ex} \sgn \omega_n}{\pi T_{cs}}}.\nonumber
\end{align}
In the $\pi$ phase state the anomalous Green's function is antisymmetric relative to $x=0$ (see Fig.~\ref{SIFS}),
\begin{equation}\label{pi}
F_f^\pi = C'(\omega_n) \sinh \left( k_f x \right).
\end{equation}
In \Eqs{zero},\eqref{pi} the $C(\omega_n)$, $C' (\omega_n)$ are the integration constants to be found from the boundary conditions.

The boundary value problem \Eqs{Usadel_S}-\eqref{S-vac} can be solved in order to obtain the closed boundary condition for $F_s$ function. At the right S/F interface ($x=d_f/2$) it acquires the form,
\begin{align}\label{boundary}
\xi_s \frac{d F_s(d_f/2)}{dx} = \frac{\gamma}{\gamma_B + B_f (\omega_n)} F_s(d_f/2).
\end{align}
Similar boundary condition can be written at $x=-d_f/2$. In \Eq{boundary} the $B_f$ function can acquire different values in 0 and $\pi$ phase states.
The zero state was already considered in Ref.~\cite{Fominov.PRB2002},
\begin{equation}
B_f^0 = \left [ k_f \xi_n \tanh (k_f d_f/2) \right]^{-1},
\end{equation}
while in the $\pi$ state from \Eq{pi} we obtain,
\begin{equation}
B_f^{\pi} = \left [ k_f \xi_n \coth (k_f d_f/2) \right]^{-1}.
\end{equation}
The boundary condition \eqref{boundary} is complex. In order to rewrite it in a real form, we use the following relation,
\begin{equation}
F^\pm = F(\omega_n) \pm F(-\omega_n).
\end{equation}
According to the Usadel equations \eqref{Usadel_S}-\eqref{Delta}, there is a symmetry relation $F(-\omega_n) = F^*(\omega_n)$ which implies that
$F^+$ is real while $F^-$ is a purely imaginary function.

Thus we can consider only positive Matsubara frequencies and express the self-consistency equation \eqref{Delta} only via the symmetric function $F^+$,
\begin{equation}\label{Delta+}
\Delta \ln \frac{T_{cs}}{T} = \pi T \sum_{\omega_n > 0} \left ( \frac{2\Delta}{\omega_n} - F_s^+ \right).
\end{equation}
The problem of $T_c$ determination can be formulated in a closed form with respect to $F_s^+$. Using the boundary condition \eqref{boundary} we arrive at the effective boundary conditions for $F_s^+$ at the right S layer boundaries,
\begin{subequations}\label{B1}
\begin{align}
\xi_s \frac{d F^+_s(d_f/2)}{dx} &= W(\omega_n) F^+_s(d_f/2),
\\
\frac{d F_s^+ (d_s+d_f/2)}{dx} &= 0,
\end{align}
\end{subequations}
Similar boundary conditions can be written at the left S layer boundaries. In \Eqs{B1} we used the notations,
\begin{align}
W^{0,\pi}(\omega_n) &= \gamma \frac{A_s \left (\gamma_B + \re B^{0,\pi}_f \right ) + \gamma}{A_s |\gamma_B + B^{0,\pi}_f|^2 + \gamma (\gamma_B + \re B^{0,\pi}_f)},
\\
A_s &= k_s \xi_s \tanh (k_s d_s), \quad k_s = \frac{1}{\xi_s} \sqrt{\frac{\omega_n}{\pi T_{cs}}}.\nonumber
\end{align}
The self-consistency equation \Eq{Delta+} and boundary conditions \Eqs{B1}, together with the Usadel equation for $F^+_s$,
\begin{equation}\label{Usadel+}
\xi_s^2 \pi T_{cs} \frac{d^2 F_s^+}{d x^2} - \omega_n F_s^+ + 2\Delta = 0
\end{equation}
can be used for finding the critical temperature of the S/F/S structure both in 0 and $\pi$ phase states. In general, this problem should be solved numerically \cite{Fominov.PRB2002}. In Ref. \cite{Fominov.PRB2002} it was also found that the so called single mode approximation (SMA) essentially simplifies the numerical problems and gives the $T_c(d_f)$ dependency with the accuracy which is enough for our consideration.
In the SMA the self-consistency equation \eqref{Delta+} takes the form\cite{Fominov.PRB2002},
\begin{equation}\label{last1}
\ln \frac{T_{cs}}{T_c} = \psi \left ( \frac{1}{2} + \frac{\Omega^2}{2}\frac{T_{cs}}{T_c} \right ) - \psi \left ( \frac{1}{2} \right ),
\end{equation}
where $\psi$ is the digamma function, and $\Omega$ can be found from the following equation,
\begin{equation}\label{last2}
\Omega \tan \left ( \Omega \frac{d_s}{\xi_s} \right ) = W^{0, \pi} (\omega_n).
\end{equation}
The critical temperature $T_c(d_f)$ is then determined by Eqs.~\eqref{last1} and \eqref{last2}. This result extends the result of Ref.~\cite{Fominov.PRB2002} to the case of S/F/S hybrid structures, where the $\pi$ phase state can be realized for large enough F layer thickness,
$d_f \sim \xi_f$.

\section{X-ray reflectometry for the determination of the layer thicknesses in periodic structures}
\begin{figure}[tb]
\centering
\includegraphics[width=\columnwidth]{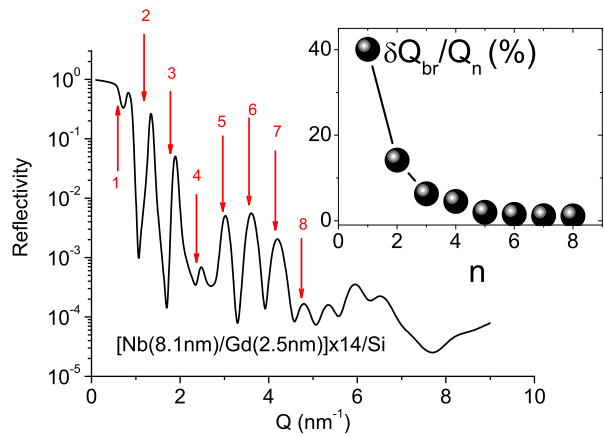}
\caption
{
Model X-ray reflectivity curve calculated for the [Nb(8.1nm)/Gd(2.5nm)]x14/Si system (compare with Fig. 1 in \cite{JiangPRL}). The vertical red arrows show the positions of the Bragg reflections calculated by Eq.~\eqref{Bragg1}. The inset shows the deviation of the peak position from the Bragg law as a function of peak order.
}
\label{Fig6}
\end{figure}

\renewcommand{\theequation}{B\arabic{equation}}
\setcounter{equation}{0}
The position of Bragg reflections in diffraction experiments from structures with periodicity $D$ can be written as

\begin{equation}\label{Bragg1}
Q_{Br}=2\pi n/D,
\end{equation}
where n is an integer. Reflectometric experiments also show Bragg-like peaks although their position  deviate from Bragg's law close to the total external reflection due to the refraction effect \cite{AndreevaPhysRevB.72.125422}. As an example we show in Fig.~\ref{Fig6} the calculated X-ray reflectivity curve for the structure [Nb(8.1nm)/Gd(2.5nm)]x14/Si described in \cite{JiangPRL}. Vertical arrows correspond to the Bragg law Eq.\eqref{Bragg1}. One can see that the positions of the maxima are shifted towards higher values comparing to the Bragg condition Eq.\eqref{Bragg1}. The deviation from the Bragg law is especially strong for smaller orders and vanishes gradually for the higher orders. Attempts to calculate the period using Eq.\eqref{Bragg1} for the n = 1 and n = 2 peaks  give D = 7.5nm and D = 9.3nm, which are 3.1nm and 1.3nm smaller than the real thickness.

\bibliography{GdNbBibliography}

\end{document}